# Transcriptional Activation of Elephant Shark Mineralocorticoid Receptor by Corticosteroids, Progesterone and Spironolactone

Yoshinao Katsu[1,2] *, Satomi Kohno[3], Kaori Oka[1], Xiaozhi Lin[1], Sumika Otake[1], Nisha E. Pillai[4], Wataru Takagi[5], Susumu Hyodo[5], Byrappa Venkatesh[4], Michael E. Baker[6], *

[1]Graduate School of Life Science, Hokkaido University, Sapporo.
[2]Department of Biological Sciences, Hokkaido University, Sapporo.
[3]Department of Biology, Saint Cloud State University, St. Cloud, MN.
[4]Comparative Genomics Laboratory, Institute of Molecular and Cell Biology, Agency for Science, Technology and Research (A*STAR), Singapore.
[5]Laboratory of Physiology, Atmosphere and Ocean Research Institute, University of Tokyo, Chiba.
[6]Division of Nephrology, Department of Medicine, University of California, San Diego, CA

*Correspondence: ykatsu@sci.hokudai.ac.jp
*Correspondence: mbaker@ucsd.edu

**Abstract:** We report the analysis of activation by corticosteroids and progesterone of full-length mineralocorticoid receptor (MR) from elephant shark, a cartilaginous fish belonging to the oldest group of jawed vertebrates. Based on their measured activities, aldosterone, cortisol, 11-deoxycorticosterone, corticosterone, 11-deoxcortisol, progesterone and 19-norprogesterone are potential physiological mineralocorticoids. However, aldosterone, the physiological mineralocorticoid in humans and other terrestrial vertebrates, is not found in cartilaginous or ray-finned fishes. Because progesterone is a precursor for corticosteroids that activate elephant shark MR, we propose that progesterone was an ancestral ligand for elephant shark MR. Although progesterone activates ray-finned fish MRs, progesterone does not activate human, amphibian or alligator MRs, suggesting that during the transition to terrestrial vertebrates, progesterone lost the ability to activate the MR. Comparison of RNA-sequence analysis of elephant shark MR with that of human MR suggests that MR expression in the human brain, heart, ovary, testis and other non-epithelial tissues evolved in cartilaginous fishes. Together, these data suggest that progesterone-activated MR may have unappreciated functions in elephant shark ovary and testis.

**Introduction**

The mineralocorticoid receptor (MR) belongs to the nuclear receptor family, a large and diverse group of transcription factors that also includes receptors for glucocorticoids (GR), progesterone (PR) androgens (AR) and estrogens (ER) (*1, 2*). Sequence analysis revealed that the MR and GR are closely related (*3*); phylogenetic analysis indicates that MR and GR evolved from a corticosteroid receptor (CR) that was present in jawless vertebrates, such as lamprey and hagfish (*4-7*). A distinct mineralocorticoid receptor (MR) first appears in cartilaginous fishes (Chondrichthyes), the oldest group of extant jawed vertebrates (gnathostomes) that diverged from bony vertebrates about 450 million years and are a crucial group in understanding the origin and evolution of jawed vertebrate morphology and physiology (*8, 9*). Like mammals, cartilaginous fishes contain the full complement of adrenal and sex steroid receptors: AR, ER, GR, MR and PR (*1, 2, 4, 10*).

Aldosterone (Aldo) is the physiological activator of transcription of human MR in epithelial tissues, such as the kidney distal collecting tubules and the colon, in which the MR regulates electrolyte homeostasis (*6, 11-14*). The MR also is found in brain, heart, aorta, lung, liver, spleen, adipose tissue, testis, breast and ovary (*12-20*), tissues in which the MR is not likely to regulate electrolyte homeostasis, its classical function. The physiological function of the MR in these tissues is still being elucidated (*14, 18, 20, 21*).

The MR and other steroid receptors have a characteristic modular structure consisting of an N-terminal domain (NTD) (domains A and B), a central DNA-binding domain (DBD) (domain C), a hinge domain (D) and a C-terminal ligand-binding domain (LBD) (domain E) (*2, 4, 22-24*) (Figure 1). The E domain alone is competent to bind steroids (*4, 22, 25-27*). Interactions between the NTD (A/B domains) and the LBD and coactivators are important regulators of transcriptional activation of mammalian MRs (*28-35*) and fish MRs (*34-36*). Moreover, GAL-DBD-hinge-LBD constructs of zebrafish MR have different responses to progesterone (Prog) and some corticosteroids than do GAL-DBD-hinge-LBD constructs of human, chicken, alligator and Xenopus MRs (*35*).

The timing of the evolution of this difference in transcriptional activation between full-length and truncated MRs in ray-finned fishes and terrestrial vertebrates, as well as when expression of the MR in non-epithelial tissues evolved is not known. Also unresolved is the identity of the ancestral mineralocorticoid in cartilaginous fishes and the current mineralocorticoid in ray-finned fish because aldosterone (Aldo), the physiological mineralocorticoid in terrestrial vertebrates, is not found in cartilaginous fishes or ray-finned fishes. Aldo first appears in lobe-finned fish (*37*), the forerunners of terrestrial vertebrates (*38*). Thus, the identity of the physiological mineralocorticoid in cartilaginous fishes and ray-finned fishes is not established, although cortisol and 11-deoxycorticosterone have been proposed (*39-45*).

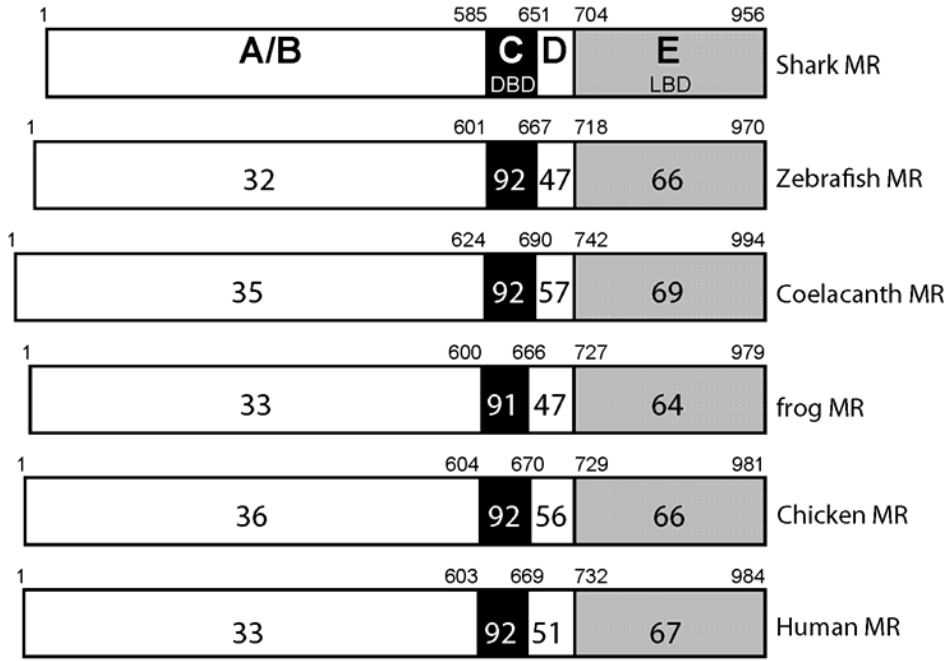

**Figure 1. Comparison of domains in elephant shark MR with vertebrate MRs.** MRs from elephant shark (shark), zebrafish, coelacanth, Xenopus (frog), chicken and human are compared. The functional A/B domain to E domains are schematically represented with the numbers of amino acid residues and the percentage of amino acid identity is depicted. GenBank accession numbers: elephant shark MR (XP_007902220), zebrafish MR (NP_001093873), coelacanth MR (XP_014348128), Xenopus (NP_001084074), chicken (ACO37437), human MR (NP_000892).

Complicating the identity of the physiological mineralocorticoid in cartilaginous and ray-finned fishes is evidence that progesterone (Prog) and 19-norprogesterone (19norProg), along with spironolactone (Spiron) (Figure 2), are transcriptional activators of several ray-finned fish MRs (*24, 36, 44*), including zebrafish MR (*35, 46*), and of chicken MR (*35*). In contrast, these steroids are antagonists for human MR (*25, 27, 47*), alligator MR and Xenopus MR (*35*). Ray-finned fish MRs and chicken MR differ in their response to Prog, 19norProg and Spiron, raising the question of whether the response to Prog and Spiron evolved in ray-finned fish before or after the divergence of ray-finned fish from the lobe-finned fish lineage that led to tetrapods.

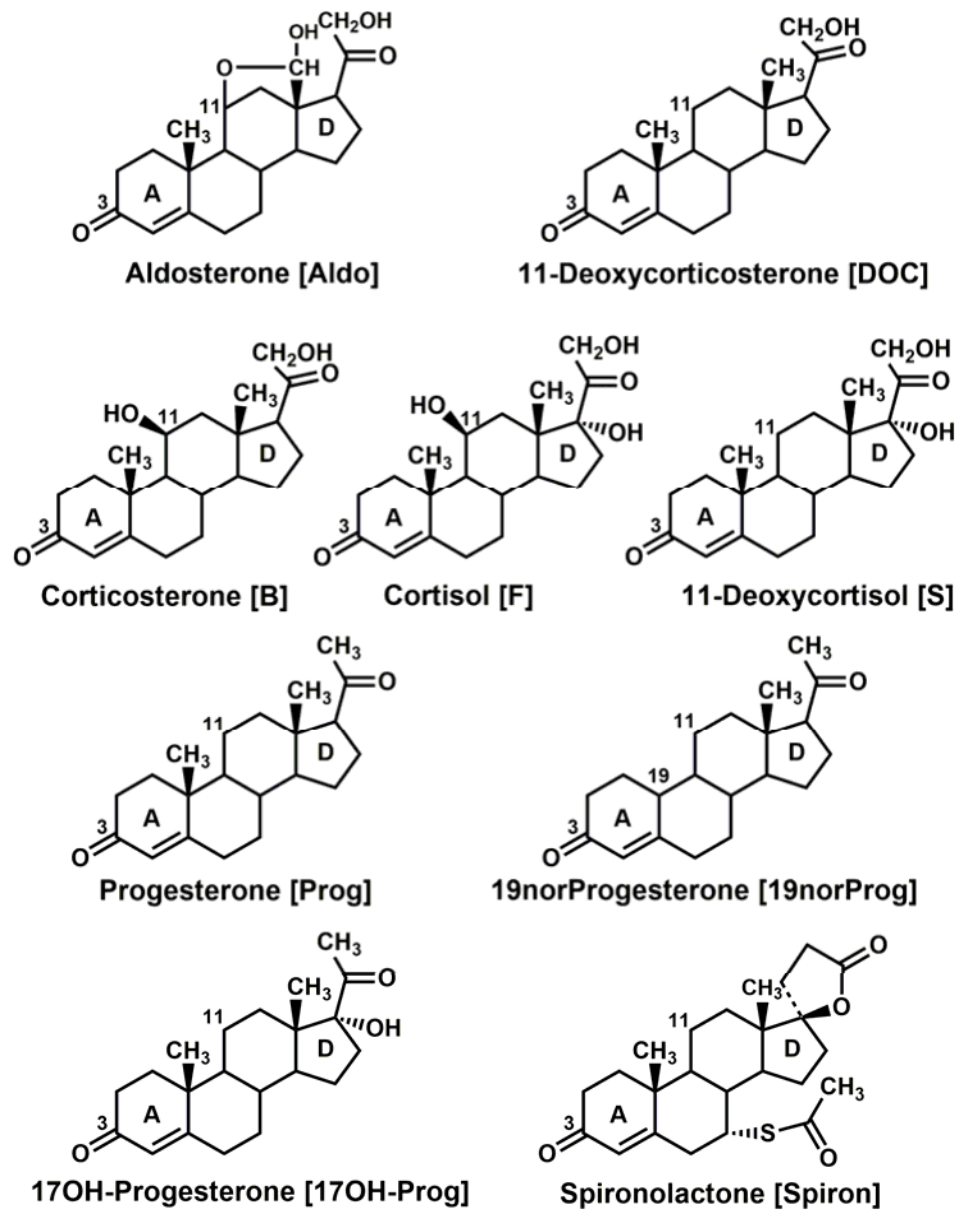

**Figure 2. Structures of steroids that are ligands for the MR.** Aldo, 11-deoxycorticosterone and S are physiological mineralocorticoids in terrestrial vertebrates (*4, 6, 12, 48*). S is both a mineralocorticoid and a glucocorticoid in lamprey (*4, 49*) and a glucocorticoid in ray-finned fish (*50*). Cortisol is a physiological glucocorticoid in terrestrial vertebrates and ray-finned fish (*4, 10, 51-53*). corticosterone is a glucocorticoid in rats and mice (*4*). Aldo, 11-deoxycorticosterone, cortisol, corticosterone and Prog have a similar high affinity for human MR (*3, 54-56*). Prog, 19norProg, 17OH-Prog and Spiron are antagonists for human MR (*24, 47, 54*) and rat MR (*57, 58*). Prog, 19norProg, and Spiron are agonists for fish MRs (*24, 36, 44*). 19norProg is a weak agonist for rat MR (*59, 60*).

Transcriptional activation by corticosteroids and other 3-ketosteroids of a full-length cartilaginous fish MR, which is the physiological MR, has not been investigated. Only truncated skate MR (*61*), consisting of the GAL4 DBD fused to the D domain and E domain of the MR

(MR-LBD) has been studied for its response to corticosteroids. In these studies, Carroll et al. found that Aldo, corticosterone, 11-deoxycorticosterone, and cortisol (Figure 2) are transcriptional activators of truncated skate MR.

Elephant shark MR is an attractive receptor to study early events in the evolution of mechanisms for regulating MR transcription because, in addition to its phylogentic position as a common ancestor of ray-finned fish and terrestrial vertebrates, genomic analyses reveal that elephant shark genes are evolving slowly (*9*), making their genes windows into the past. Thus, we investigated transcriptional activation by Aldo, 11-deoxycorticosterone, corticosterone, 11-deoxycortisol, cortisol, Prog, 19norProg, 17-hydroxyprogesterone (17OH-Prog) and Spiron of full-length and truncated elephant shark MR. Interestingly, all 3-ketosteroids, including progesterone, had EC50s of 1 nM or lower for full-length elephant shark MR. Transcriptional activation by Prog, 19norProg and Spiron of truncated elephant shark MR resembled that of zebrafish MR and not chicken MR, indicating that Prog activation of MR is an ancestral response, conserved in ray-finned fish, lost in Xenopus, alligator and human MRs, and distinct from activation of chicken MR, which arose independently. We also performed RNA-seq analysis of elephant shark MR and find widespread expression of MR in various elephant shark tissues (gill, kidney, heart, intestine, liver, spleen, brain, ovary and testis). This indicates that widespread expression of human MR in tissues, such as brain, heart, liver, spleen, ovary and testis, in which the MR does not regulate electrolyte homeostasis, evolved early in vertebrate evolution, in an ancestral cartilaginous fish. Strong MR expression in ovary and testis suggests a role for Prog-MR complexes in elephant shark reproduction, as well as other novel functions in some other MR-containing tissues. Finally, our data suggest that Prog or a related steroid may have been one of the ancestral mineralocorticoids.

**Results**
**Functional domains of elephant shark MR and other vertebrate MRs.**

In Figure 1, we compare the functional domains of elephant shark MR to selected vertebrate MRs. Elephant shark MR and human MR have 92% and 67% identity in DBD and LBD, respectively. Interestingly, elephant shark MR has similar conservation to the DBD (91-92%) and LBD (64-69%) of other MRs. The A, B and D domains of elephant shark MR and other MRs are much less conserved.

**Transcriptional activation of full-length and truncated elephant shark MR by corticosteroids, progesterone and spironolactone.**

We screened a panel of steroids at 0.1 nM and 1 nM for transcriptional activation of full-length and truncated elephant shark MR. Aldo, cortisol, corticosterone, 11-deoxycorticosterone and 11-deoxycortisol were strong activators of full-length elephant shark MR (Figure 3A) indicating that elephant shark MR has broad specificity for corticosteroids. Interestingly, at these low concentrations, Prog, 19norProg and Spiron are transcriptional activators of full-length elephant shark MR, with 19norProg having the strongest activity and 17OHProg having the weakest activity (Figure 3A).

In parallel experiments, truncated elephant shark MR, lacking the A/B domain and containing a GAL4-DBD instead of the MR DBD, retained a strong response to all corticosteroids and 19nor Prog (Figure 3B). Prog and Spiron had significant, but reduced activity, while 17OH-Prog had little activity for truncated elephant shark MR.

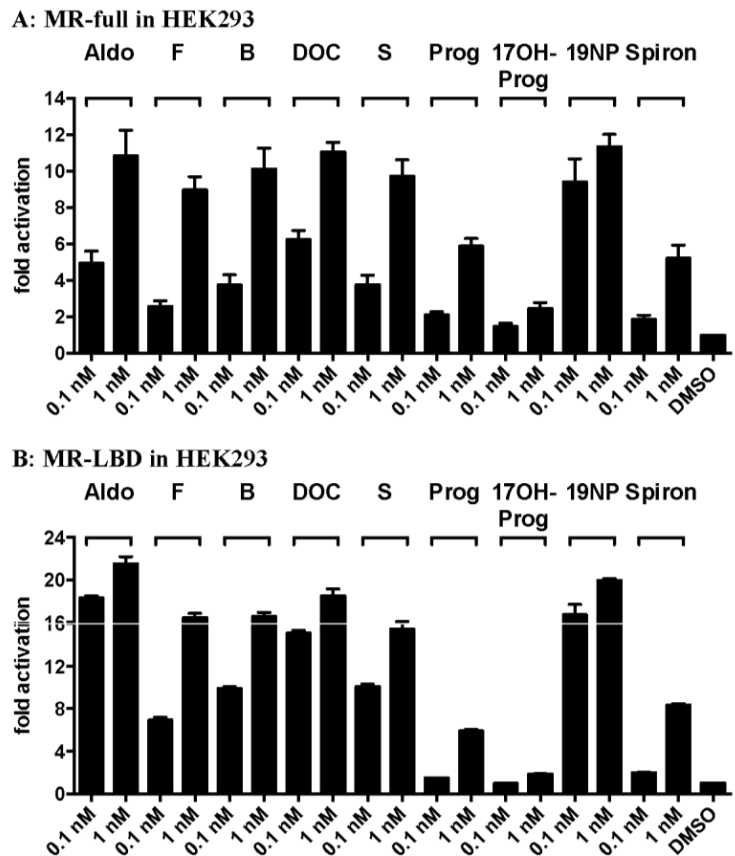

**Figure 3. Transcriptional activation of elephant shark MR by 3-ketosteroids.** Full length and truncated elephant shark MR were expressed in HEK293 cells with an MMTV-luciferase reporter.

**A.** Full length elephant shark MR. Cells were treated with 0.1 nM or 1.0 nM Aldo, cortisol, corticosterone, 11-deoxycorticosterone, 11-deoxycortisol, Prog, 19norProg (19NP), 17OH-Prog, Spiron or vehicle alone (DMSO).
**B.** Truncated elephant shark MR. Cells were treated with 0.1 nM or 1.0 nM Aldo, cortisol, corticosterone, 11-deoxycorticosterone, 11-deoxycortisol, Prog, 19norProg (19NP), 17OH-Prog or Spiron or vehicle alone (DMSO). Results are expressed as means ± SEM, n=3. Y-axis indicates fold-activation compared to the activity of control vector with vehicle (DMSO) alone as 1.

### EC50 values for steroid activation of elephant shark MR

Concentration-dependence of transcriptional activation of full length elephant shark MR by corticosteroids (Aldo, cortisol, corticosterone, 11-deoxycorticosterone, 11-deoxycortisol) is shown in Figure 4A and by Prog, 19norProg, 17OH-Prog and Spiron in Figure 4B. The corresponding concentration-dependent curves for truncated elephant shark MR are shown in Figures 4C and 4D, respectively. Table 1 summarizes the EC50s of corticosteroids for full-length and truncated elephant shark MR. Table 1 also contains, for comparison, previously determined EC50s of corticosteroids for full-length and truncated human, chicken, alligator, Xenopus and zebrafish MRs (*35*) and skate MR (*61*).

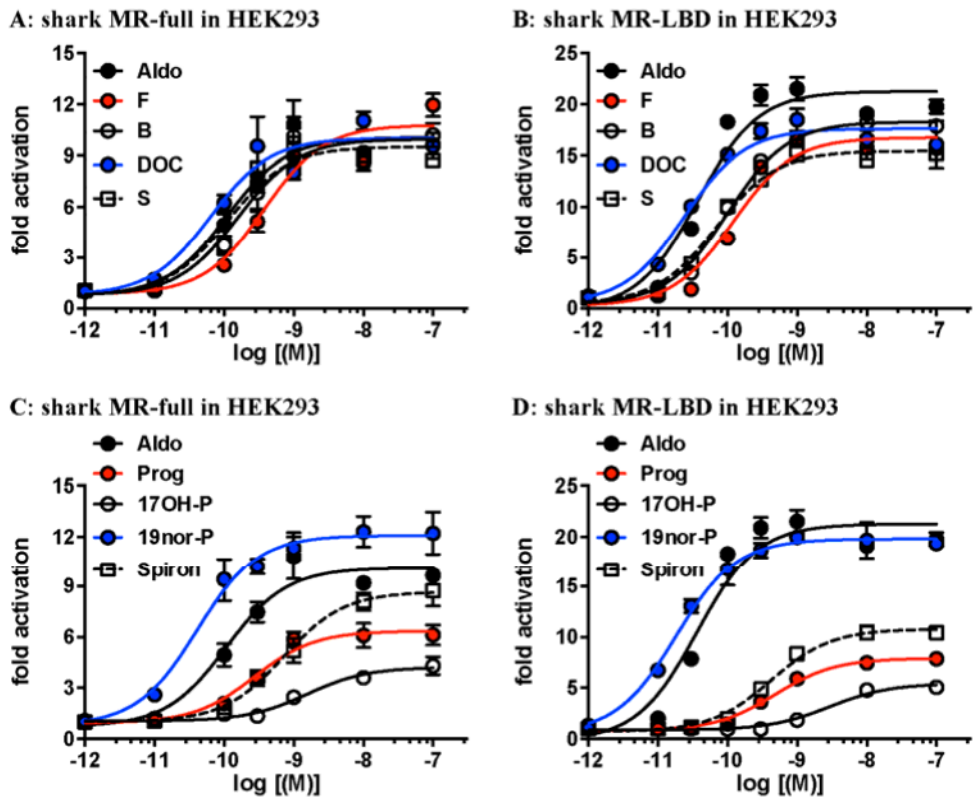

**Figure 4. Concentration-dependent transcriptional activation of full-length and truncated elephant shark MR (shark MR) by 3-ketosteroids.** Full length and truncated elephant shark MR (shark MR) were expressed in HEK293 cells with an MMTV-luciferase reporter. Full-length elephant shark MR (shark MR) (A) and (C) and truncated elephant shark MR (shark MR) (B) and (D).
**(A), (B)** Aldo, cortisol, corticosterone, 11deoxycorticosterone or 11-deoxycortisol.

**(C), (D)** Aldo, Prog, 19norProg (19nor-P), 17OH-Prog (17OH-P) or Spiron. Cells transfected with elephant shark MR were treated with increasing concentrations of cortisol, corticosterone, Aldo, 11deoxycorticosterone, 11-deoxycortisol, Prog, 19norProg, 17OH-Prog, Spiron or vehicle alone (DMSO). Y-axis indicates fold-activation compared to the activity of control vector with vehicle (DMSO) alone as 1.

**Table 1. Corticosteroid activation of full-length MR and GAL4-DBD-MR-LBD**

| MR | Aldo | B | F | DOC | S |
|---|---|---|---|---|---|
|  | EC50 (M) | EC50 (M) | EC50 (M) | EC50 (M) | EC50 (M) |
| **Elephant shark Full** | $1.1 \times 10^{-10}$ | $1.7 \times 10^{-10}$ | $4.6 \times 10^{-10}$ | $6.3 \times 10^{-11}$ | $1.1 \times 10^{-10}$ |
|  | 100% | 101% | 114% | 83% | 83% |
| **Elephant shark LBD** | $3.7 \times 10^{-11}$ | $9.9 \times 10^{-11}$ | $1.9 \times 10^{-10}$ | $2.4 \times 10^{-11}$ | $6.8 \times 10^{-11}$ |
|  | 100% | 90% | 79% | 81% | 77% |
| [1]**Skate LBD** | $7 \times 10^{-11}$ | $1 \times 10^{-10}$ | $1 \times 10^{-9}$ | $3 \times 10^{-11}$ | $2.2 \times 10^{-8}$ |
| [2]**Human Full** | $2.7 \times 10^{-10}$ | $1.2 \times 10^{-9}$ | $5.5 \times 10^{-9}$ | $4.2 \times 10^{-10}$ | $3.6 \times 10^{-9}$ |
|  | 100% | 119% | 133% | 74% | 42% |
| [2]**Human LBD** | $2.8 \times 10^{-10}$ | $5.9 \times 10^{-10}$ | $3. \times 10^{-9}$ | $1.8 \times 10^{-9}$ | #- |
|  | 100% | 95% | 74% | 44% | *8% |
| [2]**Chicken Full** | $6.2 \times 10^{-11}$ | $5.1 \times 10^{-11}$ | $2.8 \times 10^{-10}$ | $3.4 \times 10^{-11}$ | $6.7 \times 10^{-10}$ |
|  | 100% | 109% | 128% | 110% | 112% |
| [2]**Chicken LBD** | $1.3 \times 10^{-10}$ | $1.6 \times 10^{-10}$ | $6.9 \times 10^{-10}$ | $1.7 \times 10^{-10}$ | $4.7 \times 10^{-9}$ |
|  | 100% | 92% | 75% | 92% | 36% |
| [2]**Alligator-Full** | $2.8 \times 10^{-10}$ | $3.6 \times 10^{-10}$ | $6.9 \times 10^{-9}$ | $2.3 \times 10^{-10}$ | $2.7 \times 10^{-9}$ |
|  | 100% | 138% | 176% | 85% | 45% |
| [2]**Alligator LBD** | $3.5 \times 10^{-10}$ | $3.8 \times 10^{-10}$ | $2.3 \times 10^{-9}$ | $5.2 \times 10^{-10}$ | #- |
|  | 100% | 88% | 68% | 51% | *8% |
| [2]**Xenopus Full** | $4.6 \times 10^{-10}$ | $6.2 \times 10^{-10}$ | $1.1 \times 10^{-8}$ | $7.6 \times 10^{-10}$ | $9.1 \times 10^{-9}$ |
|  | 100% | 105% | 126% | 59% | 31% |
| [2]**Xenopus-LBD** | $1.5 \times 10^{-9}$ | $1.9 \times 10^{-9}$ | $1.2 \times 10^{-8}$ | #- | #- |
|  | 100% | 74% | 37% | *10% | *6% |
| [2]**Zebrafish Full** | $8.2 \times 10^{-11}$ | $3.0 \times 10^{-10}$ | $4.4 \times 10^{-10}$ | $6.3 \times 10^{-11}$ | $4.0 \times 10^{-10}$ |
|  | 100% | 112% | 123% | 103% | 94% |
| [2]**Zebrafish LBD** | $2.7 \times 10^{-11}$ | $1.5 \times 10^{-10}$ | $3.1 \times 10^{-10}$ | $1.0 \times 10^{-10}$ | $9.1 \times 10^{-10}$ |
|  | 100% | 96% | 77% | 99% | 67% |

[1] (*61*)
[2] (*35*)
\# Curve did not saturate.
(%) Relative induction to Aldosterone induced activation.
\* Relative induction at 1 μM compared to Aldo.

Previously we reported that Prog, 19norProg and Spiron are transcriptional activators of full-length chicken and zebrafish MRs and truncated zebrafish MR (*35*). However, EC50 values of Prog, 19norProg and Spiron for these MR were not determined. We have remedied this omission and report their EC50 values in Table 2 and Figure 5, for comparison with full-length and truncated elephant shark MR.

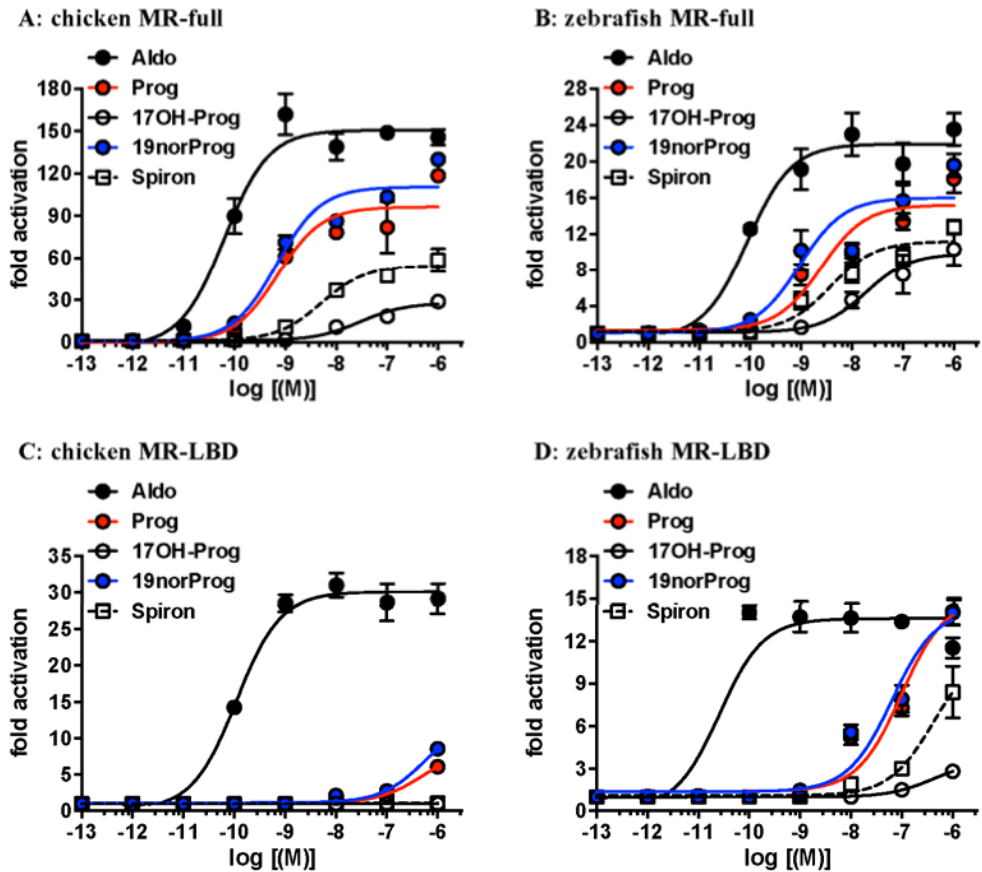

**Figure 5. Concentration-dependent transcriptional activation of full-length and truncated chicken and zebrafish MR by progestins and Spironolactone.** HEK293 cells transfected with chicken and zebrafish MR were treated with increasing concentrations of Prog, 19norProg, 17OH-Prog or Spiron. Full-length chicken MR (A) and zebrafish MR (B) and truncated chicken MR (C) and zebrafish MR (D). Y-axis indicates fold-activation compared to the activity of control vector with vehicle (DMSO) alone as 1.

**Table 2. EC50 values for progestin and spironolactone activation of full-length and GAL4-DBD-MR-LBD constructs of elephant shark, zebrafish and chicken MR**

| MR | Aldo | Prog | 17OH-Prog | 19norProg | Spiron |
|---|---|---|---|---|---|
| **Elephant shark Full** | $1.1 \times 10^{-10}$ | $2.7 \times 10^{-10}$ | $1.4 \times 10^{-9}$ | $4.3 \times 10^{-11}$ | $5.5 \times 10^{-10}$ |
| | 100% | 43% | 25% | 84% | 45% |
| **Elephant shark LBD** | $3.7 \times 10^{-11}$ | $4.8 \times 10^{-10}$ | $2.9 \times 10^{-9}$ | $1.8 \times 10^{-11}$ | $4.2 \times 10^{-10}$ |
| | 100% | 40% | 26% | 98% | 53% |
| **[2]Zebrafish Full** | $8.2 \times 10^{-11}$ | $2.4 \times 10^{-9}$ | $1.8 \times 10^{-8}$ | $9.4 \times 10^{-10}$ | $3.8 \times 10^{-9}$ |
| | 100% | 77% | 44% | 83% | 54% |
| **[2]Zebrafish LBD** | $2.7 \times 10^{-11}$ | $9.8 \times 10^{-8}$ | #- | $6.4 \times 10^{-8}$ | #- |
| | 100% | 122% | *24% | 122% | *73% |
| **[2]Chicken Full** | $6.2 \times 10^{-11}$ | $7.1 \times 10^{-10}$ | $2.9 \times 10^{-8}$ | $6.8 \times 10^{-10}$ | $5.1 \times 10^{-9}$ |
| | 100% | 62% | 15% | 68% | 30% |
| **[2]Chicken LBD** | $1.3 \times 10^{-10}$ | #- | #- | #- | #- |
| | 100% | *21% | - | *29% | - |

[2] (*35*) # Curve did not saturate.
(%) Relative induction to Aldosterone induced activation.
* Relative induction at 1 μM compared to Aldo.

**RNA-seq analysis of elephant shark MR**

We examined expression of level of elephant shark MR gene in 10 tissues based on RNA-seq data (Figure 6A). The MR gene was found to express widely in all tissues, including gills and kidney, two traditional mineralocorticoid-responsive tissues. Interestingly there was considerably higher expression in the ovary and testis, the two reproductive tissues analyzed.

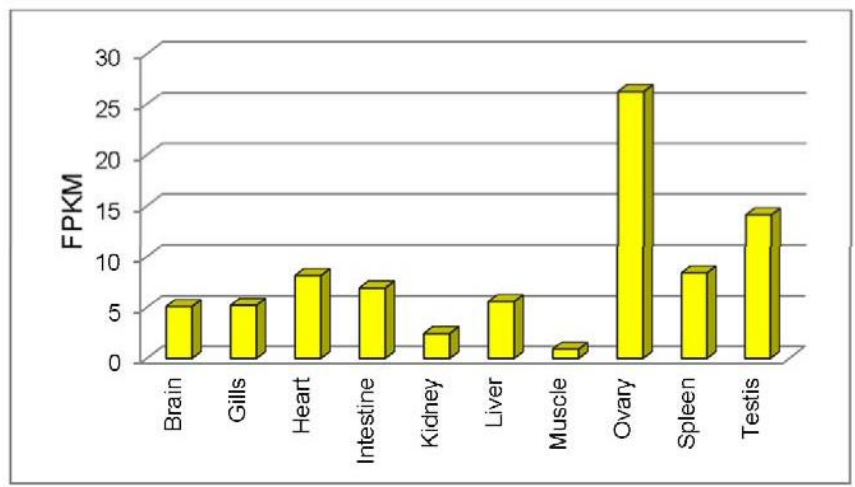

**Figure 6A. Expression level of MR gene in 10 tissues of elephant shark estimated based on RNA-seq data.** Transcript abundances are shown in terms of normalized counts called Fragments per kilobase of exon per million fragments mapped (FPKM) (*62*). FPKM is estimated by normalizing gene length followed by normalizing for sequencing depth.

**RNA-seq analysis of human MR**

RNA-seq analysis of human MR (*63*) (Figure 6B) reveals that the MR is expressed in kidney, colon, brain, heart, liver, ovary, spleen and testis. This diverse pattern of expression is similar to that of elephant shark MR.

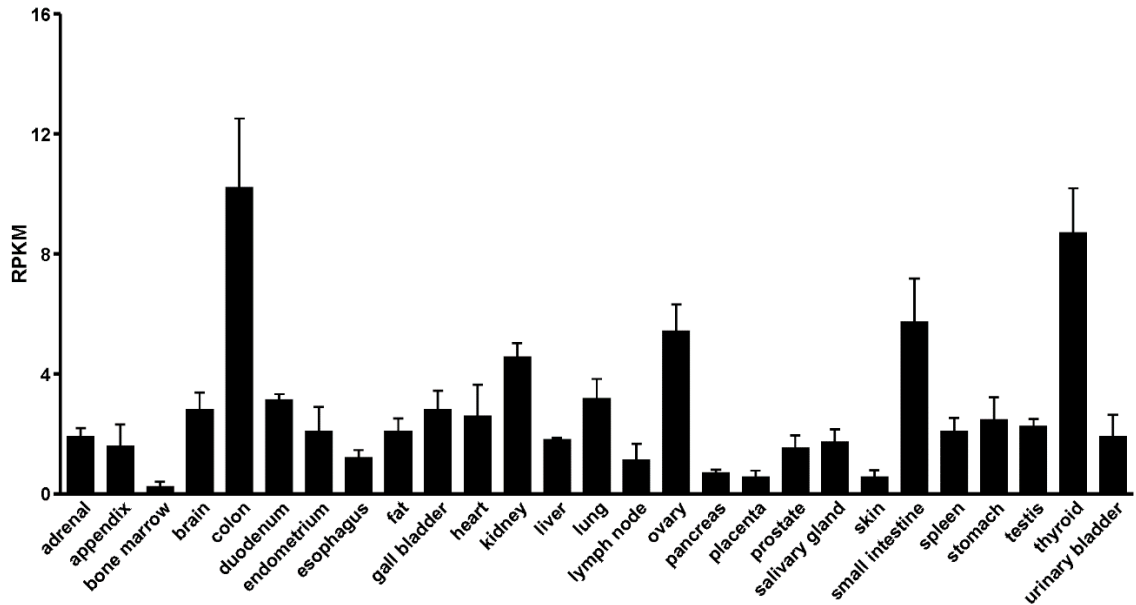

**Figure 6B. Expression level of human MR based on RNA-seq data.** Transcript abundances are shown in terms of normalized counts called Reads Per Kilobase of transcript per Million mapped reads (RPKM) (*63*).

**Discussion**

Cartilaginous fish, including elephant sharks, occupy a key position in the evolution of vertebrates as an out-group to ray-finned fish, the largest group of extant vertebrates, and the lobe-finned fish, which are the forerunners of terrestrial vertebrates. Importantly, the elephant shark genome is evolving slowly (*9*), making it attractive for studying ancestral proteins, including the MR, which first appears as a distinct MR ortholog in cartilaginous fish (*5, 6, 10, 61*).

This is the first report of transcriptional activation of a full-length cartilaginous fish MR by a panel of five corticosteroids and Prog, 19norProg, 17OH-Prog and Spiron. Full-length elephant shark MR has EC50s below 1 nM for five corticosteroids, (Figure 4, Table 1) including Aldo, a steroid that has not been found in cartilaginous fish. Prog, 19norProg and Spiron also have sub-nM EC50s for full-length elephant shark MR. In addition to their low EC50s, all of these corticosteroids and 19norProg have similar fold activation of transcription of full-length MR (Figure 4). Thus, several corticosteroids, as well as 19norProg and Prog, are potential physiological mineralocorticoids for elephant shark MR.

Compared to full-length elephant shark MR, the EC50s of all five corticosteroids and 19norProg are lower for its truncated MR, while the EC50 for Spiron is slightly lower, and the

EC50s for Prog and 17OH-Prog are about 2-fold higher (Table 1, Table 2).  Compared to fold activation by Aldo, the responses to 11-deoxycorticosterone, S, Prog and 17OH-Prog are similar for full-length and truncated elephant shark MR.  However, the relative responses to corticosterone and cortisol of truncated elephant shark MR are 10% and 35% lower, respectively, (Figure 4, Table 1) and for 19norProg and Spiron, 14% and 8% higher respectively, than for full-length elephant shark MR (Figure 4, Table 2).  Regarding truncated skate MR, most of the EC50s of corticosteroids (*61*) are similar to that for elephant shark MR (Table 1).  The exception is 11-deoxycortisol, which has over 200-fold higher EC50 for skate MR than for elephant shark MR.

**Corticosteroid Progesterone and 19norProgesterone activation of full-length and truncated elephant shark, terrestrial vertebrate and zebrafish MRs**

Comparison of transcriptional activation by corticosteroids, Prog and 19norProg of full-length and truncated elephant shark MR with full-length and truncated human, chicken, alligator, Xenopus and zebrafish MRs (Table 1) provides insights into the evolution of regulation of steroid-mediated transcription of these MRs by allosteric interactions between the NTD/DBD and LBD.  Like elephant shark MR, truncated terrestrial vertebrate MRs and zebrafish MR have similar EC50s for Aldo, corticosterone and cortisol as their full-length counterparts (Table 1). However, the response of truncated human, alligator and Xenopus MRs to 11-deoxycortisol did not saturate at 1 µM, preventing us from determining the EC50 and explaining the low fold activation by S (Table 1).  Interestingly, truncated chicken and zebrafish MRs, which are activated by Prog and19norProg, have nM EC50s for 11-deoxycortisol, although fold activation by 11-deoxycortisol is lower than that of Aldo.  The similar activity of truncated human, alligator and Xenopus MRs to Aldo, corticosterone and cortisol indicates that allosteric interactions between the NTD and LBD are steroid-selective.  Overall, it appears that corticosteroid activation, especially by 11-deoxycortisol, of elephant shark MR is less sensitive to the loss of NTD/DBD than are human, chicken, alligator and Xenopus MRs.

**Progesterone and 19norProgesterone activation of full-length and truncated chicken and zebrafish MRs**

Transcriptional activation by Prog and 19norProg of chicken and zebrafish MRs differs from that of elephant shark MR, especially for their truncated MRs.  Although Prog, 19norProg and Spiron have nM EC50s for transcriptional activation of full-length chicken and zebrafish

MRs, at 1 µM Prog 19norProg and Spiron transcriptional activation of truncated chicken MR did not saturate, and at 1 µM Spiron transcriptional activation of zebrafish MR did not saturate (Figure 5, Table 2). Fold activation compared to Aldo by Prog, 19norProg and Spiron is substantially lower for truncated chicken MR and for Spiron activation of zebrafish MR. However, Prog and 19norProg have EC50s of 98 nM and 64 nM, respectively, for truncated zebrafish MR. In this respect, truncated zebrafish MR is closer to truncated elephant shark MR than is truncated chicken MR. These data suggest that transcriptional activation by Prog and 19norProg of chicken MR appears to have evolved independently in terrestrial vertebrates.

**Progesterone may be a mineralocorticoid in cartilaginous fishes**

Prog is a precursor for the synthesis of the other corticosteroids (*5, 64*). Two parsimonious metabolites of progesterone are 11-deoxycorticosterone (21-hydroxyprogesterone) and 19norProg (Figure 2), which have EC50s, 0.063 nM and 0.043 nM, respectively, for elephant shark MR. These steroids have the two lowest EC50s among the tested steroids. Moreover, 19norProg evokes a stronger response from elephant shark MR than Aldo (Figure 4A, B). C19 demethylase activity has been found in mammalian kidney (*59*). If C19 demethylase is present in elephant shark, then 19norProg needs to be considered as a potential physiological mineralocorticoid.

The strong response to 19norProg is interesting because *in vivo* studies in rats revealed that 19norProg is at least a 100-fold weaker MR agonist than Aldo (*59*), while in transfected cells, at 1 nM, 19norProg is an antagonist (*24, 47*). Unexpectedly, compared to Aldo, 19norAldo has less than 1% binding affinity for rat MR (*65*) indicating that the mechanism by which C19 demethylation of Prog increases this steroid's transcriptional activity for elephant shark MR is likely to be complex.

We propose that transcriptional activation of elephant shark MR by 19norProg, as well as by Prog and Spiron, can be explained by a mechanism based on Geller et al.'s (*47*) discovery that S810L human MR mutant is transcriptionally activated by 1 nM Prog, 19norProg, and Spiron, unlike wild-type human MR, in which these steroids are MR antagonists. Geller et al. used a 3D model of S810L MR and transcriptional analysis of a series of mutations at Ser-810 (helix 5) and Ala-773 (helix 3) to propose that a van der Waals contact between Leu-810 and Ala-773 was sufficient for transcriptional activation of S810L MR by Prog and 19norProg. Figure 3C of Geller et al. (*47*) shows that the human S810M mutant was activated by 19norProg. Elephant shark MR and skate MR contain a methionine at this position and an alanine corresponding to Ala-773 (Figure 7). Based on Geller et al.'s model, we propose that transcriptional activation of elephant shark MR by 19norProg is due to a van der Waals contact between Met-782 (helix 5) and Ala-745 (helix 3), which stabilizes the A ring of 19norProg, promoting transcriptional activation.

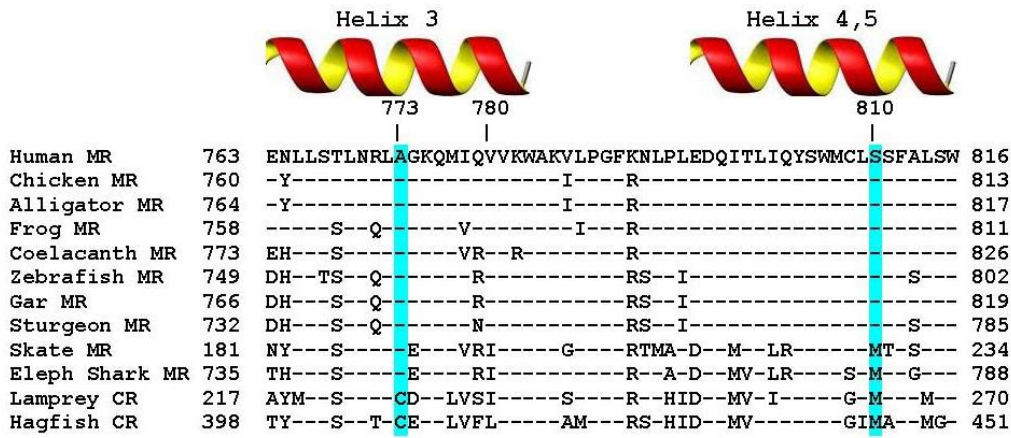

**Figure 7. Alignment of elephant shark MR to Serine-810 and Alanine-773 in helices 3-5 in human MR.** Elephant shark MR and skate MR contain a methionine corresponding to human Ser-810 and an alanine corresponding to Ala-773. Lamprey CR and hagfish CR also contain a corresponding methionine, as well as a cysteine corresponding to Ala-773. Human Ser-810 and Ala-773 are conserved in coelacanths, terrestrial vertebrate and ray-finned fish MRs. Amino acids that are identical to amino acids in human MR are denoted by (-).

Our results indicate that Prog activation of the MR is an ancient response. The timing of evolution of Prog as an antagonist for the MR is not known. One necessary event appears to be the evolution in helix 5 of a serine corresponding to serine-810 in human MR. This occurs in coelacanth MR, which contains Ser-816, corresponding to human MR Ser-810 (Figure 7) and to elephant shark MR Met-782. This mutation in coelacanth MR would be expected to eliminate the proposed van der Waals contact between helix 5 and helix 3, leading to antagonist activity of Prog in some terrestrial vertebrate MRs (*35, 47*). However, a serine corresponding to Ser-810 in human MR also is present in chicken MR and ray-finned fish MRs (Figure 7), indicating another mechanism is involved in Prog activation of the MR in these vertebrates.

**Role for elephant shark MR in reproductive physiology**

The evidence from RNA-seq analysis of high MR expression in elephant shark ovary and testis (Figure 6A) suggests that a Prog-MR complex is important in reproductive responses in elephant shark. Of course, Prog also acts as a reproductive steroid in ovary and testis via transcriptonal activation of the PR (*66, 67*).  Based on evidence that Prog activates the MR in several ray-finned fishes (*24, 36, 44, 46*), a Prog-MR complex also may be active in reproductive tissues and other tissues in ray-finned fish, as well as in cartilaginous fish.

RNA-seq analysis also finds MR expression in elephant shark gills and kidneys (Figure 6A), two classical targets for MR regulation of electrolyte transport (*6*).  Moreover, RNA-seq analysis also identifies MR expression in elephant shark heart and brain, two other tissues in which corticosteroids have important physiological actions via the MR (*16-18, 20, 21, 68-70*).  Indeed, RNA seq analysis of elephant shark MR indicates that expression the MR in diverse tissues was conserved during the descent from cartilaginous fishes to humans.  Expression of shark MR in many tissues (brain, heart, liver, ovary) in which the MR is not likely to regulate electrolyte homeostasis, the MR's classical function, further supports evidence from the last thirty years (*3, 17-19, 68, 70-72*) that mineralocorticoid activity is an incomplete functional description of this nuclear receptor.  An alternative name is needed to describe more completely the functions of the MR.

## Materials and Methods
### Chemical reagents
Aldo, F, B, DOC, S, Prog, 19norProg, 17OH-Prog and Spiron were purchased from Sigma-Aldrich.  For reporter gene assays, all hormones were dissolved in dimethylsulfoxide (DMSO); the final DMSO concentration in the culture medium did not exceed 0.1%.

### Construction of plasmid vectors
The full-coding regions from elephant shark MR were amplified by PCR with KOD DNA polymerase.  The PCR products were gel-purified and ligated into pcDNA3.1 vector (Invitrogen) for the full-coding region or pBIND vector for D-E domains (*73*).

### RNA-sequence analysis
The RNA-seq reads from following tissues of elephant shark were down-loaded from the Sequence Read Archive database of NCBI (accession number SRA054255): brain, gills, heart, intestine, kidney, liver, muscle, ovary, spleen, and testis.  For each tissue, sequences were assembled *de novo* using Trinity, version r2013-08-14 (*74*).

### Gene Expression Analyses
To determine the expression level of MR genes, we performed abundance estimation of transcripts from the afore mentioned 10 tissues. Trinity transcripts from all ten tissues and full-length cDNA sequence of the MR gene were combined together and clustered using CD-HITv4.6.1 at 100% identity (*75*). RNA-seq reads from each of the ten tissues was independently aligned to the clustered transcript sequences and abundance of MR gene transcripts was estimated by RSEMv1.2.25 (*62*) which uses bowtiev2.2.6 for aligning (*76*). Transcript abundances were measured in terms of normalized counts called Fragments per kilobase of exon per million fragments mapped (FPKM) (*62*). FPKM is estimated by normalizing the gene length followed by normalizing for sequencing depth.

**Transactivation Assay and Statistical Methods**

Transfection and reporter assays were carried out in HEK293 cells, as described previously (*73, 77*). All experiments were performed in triplicate. The values shown are mean ± SEM from three separate experiments, and dose-response data and EC50 were analyzed using GraphPad Prism. Comparisons between two groups were performed using *t*-test, and all multi-group comparisons were performed using one-way ANOVA followed by Bonferroni test. $P < 0.05$ was considered statistically significant.

**Acknowledgments:**
**Funding**: K.O. was supported by the Japan Society for the Promotion of Science (JSPS) Research Fellowships for Young Scientists. This work was supported in part by Grants-in-Aid for Scientific Research 23570067 and 26440159 (YK) from the Ministry of Education, Culture, Sports, Science and Technology of Japan. M.E.B. was supported by Research fund #3096. BV was supported by the Biomedical Research Council of A*STAR, Singapore.
**Author contributions:** Y.K., S.K., K.O., X.L., S.O. and N.E.P. carried out the research. Y.K. and S.K. analyzed data. W.T. and S.H. aided in the collection of animals. NEP assembled and analyzed RNAseq data. Y.K. and M.E.B. conceived and designed the experiments. Y.K., M.E.B. and B.V. wrote the paper. All authors gave final approval for publication.
**Declaration of interests**: We have no competing interests.